# FPGA Based Implementation of Distributed Minority and Majority Voting Based Redundancy for Mission and Safety-Critical Applications

P. BALASUBRAMANIAN*, N.E. MASTORAKIS[§¶]
* School of Computer Engineering
Nanyang Technological University
50 Nanyang Avenue
Singapore 639798
E-mail: balasubramanian@ntu.edu.sg
[§] Department of Computer Science
Military Institutes of University Education
Hellenic Naval Academy
Piraeus 18539, Greece
E-mail: mastor@hna.gr
[¶] Department of Industrial Engineering
Technical University of Sofia
Sofia 1000, Boulevard Kliment Ohridski 8
Bulgaria
E-mail: mastor@tu-sofia.bg

*Abstract:* - Electronic circuits and systems used in mission and safety-critical applications usually employ redundancy in the design to overcome arbitrary fault(s) or failure(s) and guarantee the correct operation. In this context, the distributed minority and majority voting based redundancy (DMMR) scheme forms an efficient alternative to the conventional N-modular redundancy (NMR) scheme for implementing mission and safety-critical circuits and systems by significantly minimizing their weight and design cost and also their design metrics whilst providing a similar degree of fault tolerance. This article presents the first FPGAs based implementation of example DMMR circuits and compares it with counterpart NMR circuits on the basis of area occupancy and critical path delay viz. area-delay product (ADP). The example DMMR circuits and counterpart NMR circuits are able to accommodate the faulty or failure states of 2, 3 and 4 function modules. For physical synthesis, two commercial Xilinx FPGAs viz. Spartan 3E and Virtex 5 corresponding to 90nm and 65nm CMOS processes, and two radiation-tolerant and military grade Xilinx FPGAs viz. QPro Virtex 2 and QPro Virtex E corresponding to 150nm and 180nm CMOS processes were considered for the NMR and DMMR circuit realizations which employ the 4×4 array multiplier as a representative function module. To achieve a fault tolerance of 2 function modules, both the DMMR and the NMR schemes provide near similar mean ADPs across all the four FPGAs. But while achieving a fault tolerance of 3 function modules the DMMR features reduced ADP by 44.5% on average compared to the NMR, and in achieving a fault tolerance of 4 function modules the DMMR reports reduced ADP by 56.5% on average compared to the NMR with respect to all the four FPGAs considered.

## 1 Introduction
Mission and safety-critical circuits and systems used in niche applications such as space, aerospace, defense, nuclear, banking and finance such as commercial banking systems and stock exchanges and other applications such as power systems, industrial automation and control etc. inherently employ redundancy in the design to cope with arbitrary function module fault(s) or failure(s) and still guarantee the correct operation [1]. Here the term 'function module' refers to any arbitrary electronic circuit or system.

In a typical passive N-modular redundant design, (N–1) identical copies of a function module are used along with the primary function module, and at least a majority (N+1)/2 out of the N function modules should always maintain the correct operation to guarantee the correct operation of the NMR design [2]. Therefore the faulty or failure state(s) of at most (N–1)/2 function modules is tolerated by the NMR design. The respective output(s) of the N function modules are combined using voting element(s) which perform majority voting on the function modules output(s) and generate the primary outputs.

Although the NMR scheme is well established, widely understood and used, it tends to be unsuitable for coping with more than 2 function module faults or failures [3]. When higher fault tolerances are demanded by a mission or safety-critical circuit or system entirely or selectively, the NMR scheme may not be preferable since it would exacerbate the design metrics and increase the design weight and cost due to the requirement for provision of more identical copies of the function module. This in fact assumes significance in the light of the observation [4] [5] that multiple faults are imminent and are likely to become more common in nanometer scale electronic designs deployed in mission and safety-critical applications owing to the adverse impact of radiation phenomena on small device geometries. In order to overcome these, higher levels of redundancy may be inevitable at least selectively in a mission or safety-critical circuit or system design [6], i.e. higher levels of redundancy may have to be implemented selectively in at least the sensitive portions of a mission or safety-critical electronic circuit or system.

To mitigate the excessive design overheads associated with higher order NMR designs whilst providing similar degree(s) of fault tolerance the DMMR scheme was proposed [7]. In a DMMR scheme, supposing M identical function modules are considered, they are split into two groups as 3 function modules constituting the majority logic group and the remaining (M–3) function modules constituting the minority logic group in which case the DMMR is labelled as the 3-of-M DMMR. Supposing 5 function modules are deployed in the majority logic group and (M–5) function modules are deployed in the minority logic group, the DMMR system is referred to as a 5-of-M DMMR system. It was recently shown in [8] that the 3-of-M DMMR system architecture is preferable over the 5-of-M DMMR system architecture overall in terms of system reliability, fault tolerance and the design metrics.

In the 3-of-M DMMR scheme, in the majority logic group, the faulty or failure state of any arbitrary function module is tolerated, and in the minority logic group a minimum of 1 out of the (M–3) function modules should maintain the correct operation. Thus the minority logic group can easily mask the faulty or failure state(s) of utmost (M–4) function modules. The 3-of-M DMMR design, as a minimum, should incorporate 5 identical function modules (i.e. M = 5) with 3 identical function modules constituting the majority logic group and the remaining 2 function modules constituting the minority logic group called the 3-of-5 DMMR design.

The biggest advantage of the (3-of-M) DMMR scheme is that with the introduction of every extra function module in the minority logic group the fault tolerance of the DMMR scheme proportionately increases by unity. It is worth noting here that 2 function modules have to be added to an NMR design to increase its fault tolerance by unity while only 1 function module has to be added to a (3-of-M) DMMR design to enhance its fault tolerance by unity [7]. This is indeed beneficial given that without any compromise on the fault tolerance, the design weight and cost could be significantly reduced in the DMMR scheme compared to the conventional NMR scheme due to the usage of less function module(s), and thus the design metrics would be better optimized in the case of the former compared to the latter.

This article for the first time presents the FPGAs based implementation of example (3-of-M) DMMR designs and compares it with counterpart NMR designs for fault or failure tolerances of 2, 3 and 4 function modules. The rest of this article is organized into 3 sections. Section 2 briefly discusses the NMR and DMMR system architectures. Section 3 describes example implementations of NMR and DMMR circuits targeting four different FPGAs viz. two commercial FPGAs (Spartan 3E and Virtex 5), a radiation-tolerant FPGA (QPro Virtex 2) and a military grade FPGA (QPro Virtex E), and presents the synthesis results obtained viz. area and delay combined into the area-delay product (i.e. ADP). Lastly Section 4 gives the conclusions.

## 2 Description of NMR and DMMR Schemes

The generic architectures of the NMR and DMMR schemes are succinctly discussed in this section.

### 2.1 NMR Scheme

The general architecture of the NMR scheme is depicted through Figure 1. A similar set of inputs is supplied to all the N identical function modules from the external environment. In Figure 1, the outputs of N identical function modules viz. $M_1$ to $M_N$ of NMR are combined using a majority voter which produces the NMR design output (NMRO) after performing majority voting on the function modules outputs. At the least (N+1)/2 out of the N identical function modules should operate correctly to withstand the faulty or failure state(s) of a maximum of (N–1)/2 function modules.

P. Balasubramanian, N.e. Mastorakis

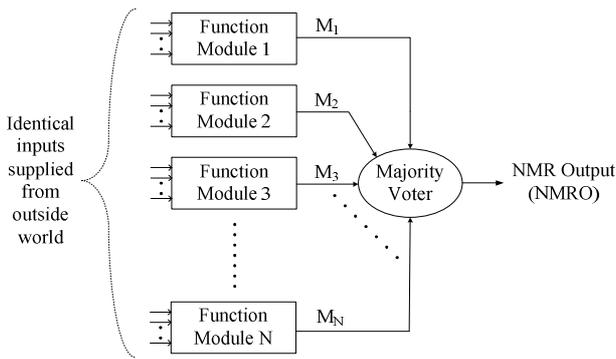

Fig. 1 Block schematic of NMR scheme

The basic NMR scheme corresponds to the 3MR scheme (i.e. triple modular redundancy or TMR) where 3 identical function modules are used and the faulty or failure state of at most 1 function module is tolerated. 5MR, wherein 5 identical function modules are used would guarantee the correct operation provided at most 2 function modules may become faulty or fail at random. In 7MR, 7 identical function modules are used and at least 4 function modules should operate correctly thus being able to accommodate the faulty or failure state of at most 3 function modules. 9 identical function modules are used in the 9MR and a minimum of 5 function modules must maintain the correct operation to successfully mask the faulty or failure state of a maximum of 4 function modules. Hence it becomes clear that according to the NMR scheme, 2 identical function modules have to be added to a NMR design in order to enhance its fault tolerance by unity. This poses major drawbacks in terms of exaggerating the design metrics and substantially increasing the design weight and cost. Moreover the voters' complexity of the NMR scheme [7] also increases considerably with an increase in the NMR design hierarchy.

## 2.2 DMMR Scheme

The general DMMR architecture is shown in Figure 2 that consists of M identical function modules which are split into two groups as the 'majority logic group' comprising the function modules $F_1$, $F_2$ and $F_3$, and the 'minority logic group' comprising the function modules $F_4$ to $F_M$. The majority and minority logic groups are shown enclosed in brown and blue rectangles respectively in Figure 2. The DMMR voter is highlighted by the pink rectangle in Figure 2. The majority voter, which forms part of the DMMR voter, performs majority voting on only the majority logic group function modules outputs viz. $F_1$, $F_2$ and $F_3$ and produces the intermediate output MAJ. The gate level detail of the 3-input majority voter [9] is shown in dotted lines in Figure 2, and is similar to the carry output logic of the binary full adder [12]. The outputs of the minority logic group function modules viz. $F_4$ to $F_M$ are combined using an OR gate whose output is MIN. The logical conjunction of MAJ and MIN yields the DMMR design output i.e. DMMRO. The logic equations of MAJ, MIN and DMMRO are given below. In (1), (2) and (3), '+' signifies logical disjunction and '●' or the product signifies logical conjunction.

$$MAJ = F_1F_2 + F_2F_3 + F_1F_3 \qquad (1)$$

$$MIN = F_4 + F_5 +\ldots+ F_M \qquad (2)$$

$$DMMRO = MAJ \bullet MIN \qquad (3)$$

To briefly discuss the DMMR design architecture, let us first assume that the correct steady-state of all the function module outputs viz. $F_1$ to $F_M$ should be binary 1 in Figure 2. Supposing due to faults or failures of function module 3 and function modules 5 to M, let us presume their outputs are corrupted. As a result, $F_3$ and $F_5$ up to $F_M$ assume binary 0. Hence, as per our assumptions, $F_1 = F_2 = F_4 = 1$. Since $F_1$ and $F_2$ are 1, as per (1), MAJ would evaluate to 1. Since $F_4 = 1$, MIN equates to 1 as per (2) although $F_5$ up to $F_M$ have incorrectly assumed binary 0 due to faults or failures. Since MAJ and MIN are 1, as per (3) the DMMR architecture outputs 1 on DMMRO which is correct.

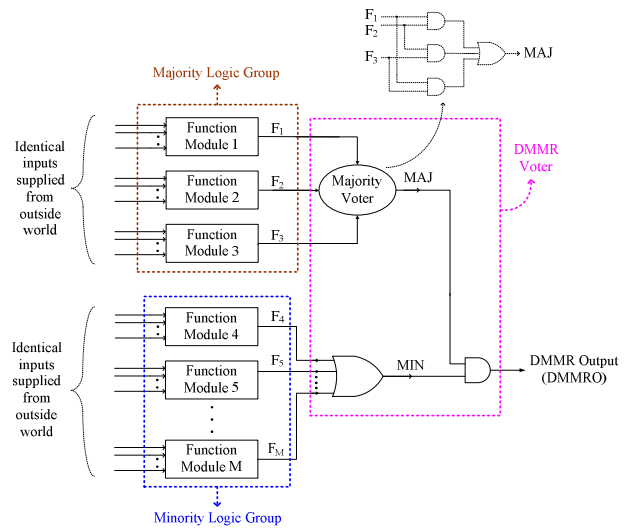

Fig. 2 Block schematic of (3-of-M) DMMR scheme

On similar lines, when all the function module outputs say $F_1$ to $F_M$ should be binary 0 and if only $F_1$, $F_2$ and $F_4$ are binary 0 and the rest of the outputs viz. $F_3$ and $F_5$ up to $F_M$ assume binary 1 incorrectly, as per (1) and (2), we find that MAJ would evaluate to binary 0 which is correct but MIN would evaluate to binary 1 which is incorrect. Nevertheless, as per (3), DMMRO would correctly equate to binary 0 implying that the DMMR system architecture is able to successfully mask the faulty or failure states of various function modules and could guarantee the correct operation provided the Boolean majority and minority logic conditions are simultaneously upheld in the corresponding majority and minority logic groups of the function modules shown in Figure 2.

The majority logic group can withstand the failure or faulty state of anyone of the 3 function modules among $F_1$, $F_2$ and $F_3$. The minority logic group is more accommodative and can withstand the faulty or failure state(s) of all but one of the function modules among $F_4$ to $F_M$. This implies that the introduction of each extra function module in the minority logic group increases the fault tolerance of the DMMR scheme by unity. This is advantageous since the NMR scheme requires the addition of 2 function modules to improve its fault tolerance by unity. Hence, given this, the DMMR scheme could help in reducing the number of function modules used compared to the NMR scheme in order to achieve the same degree of fault tolerance whilst being able to reduce the design weight and cost and also help in optimizing the design metrics.

## 3 Example FPGA Based Realizations of NMR and DMMR Circuits – Results and Discussion

Two commercial Xilinx FPGA families viz. Spartan 3E (Device: XC3S1600E) and Virtex 5 (Device: XC5VLX30T) corresponding to 90nm and 65nm CMOS processes, a radiation-tolerant FPGA family viz. QPro Virtex 2 (Device: XQR2V1000), and a military grade FPGA family viz. QPro Virtex E (Device: XQV600E) corresponding to 150nm and 180nm CMOS processes have been considered as the FPGA implementation platforms. The radiation-tolerant and military grade FPGA families are particularly considered since the NMR and DMMR schemes are generally suitable for deployment in mission and safety-critical applications. A 4×4 Braun array multiplier [10] portrayed by Figure 3 has been considered as the representative function module although any function module representing any electronic circuit or system could be considered for the NMR or DMMR designs depending upon the target application.

FPGAs based implementations of example 5MR, 7MR and 9MR circuits, and their respective redundant counterparts viz. 3-of-5 DMMR, 3-of-6 DMMR and 3-of-7 DMMR circuits as per the generic NMR and (3-of-M) DMMR architectures shown in Figures 1 and 2 were considered for comparison. The 5MR, 7MR and 9MR circuits correspond to the generic NMR scheme, and the 3-of-5 DMMR, 3-of-6 DMMR and 3-of-7 DMMR circuits correspond to the generic (3-of-M) DMMR scheme. The redundant circuits comprise identical function modules and different voter circuits. The voters corresponding to various NMR and (3-of-M) DMMR schemes were implemented according to the gate-level schematics given in [9] [11].

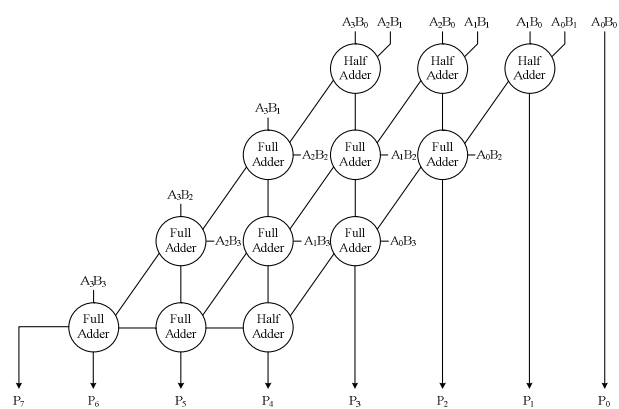

Fig. 3 Logic schematic of 4×4 array multiplier

In the 3-of-5, 3-of-6 and 3-of-7 DMMR circuits, 3 function modules constitute their majority logic groups while their respective minority logic groups comprise 2, 3 and 4 function modules each. The 5MR and 3-of-5 DMMR circuits both consist of 5 function modules and both these provide maximum fault or failure tolerance capability of 2 function modules. The 7MR circuit comprises 7 function modules while the 3-of-6 DMMR circuit consists of only 6 function modules. However the 3-of-6 DMMR circuit despite requiring 1 function module less than the 7MR circuit features a similar fault or failure tolerance capability of maximum of 2 function modules as its counterpart. The 9MR circuit consists of 9 function modules while the 3-of-7 DMMR circuit consists of just 7 function modules. The 3-of-7 DMMR circuit despite requiring 2 function modules less than the 9MR circuit exhibits a similar fault or failure tolerance capability of maximum of 4 function modules as its counterpart. These mean the area occupancy, critical path delay and ADP metrics of the 3-of-6 DMMR and 3-of-7 DMMR circuits would be favorably optimized i.e. less than the corresponding design parameters of 7MR and 9MR circuits as substantiated by the results given in Table 1.

Table 1 gives the simulation results viz. area in terms of the number of basic logic elements (BELs)

and the critical path delay corresponding to 5MR, 7MR, 9MR, 3-of-5 DMMR, 3-of-6 DMMR and 3-of-7 DMMR circuits based on the four different FPGAs considered. The corresponding ADP values are also mentioned in the last column of Table 1.

Table 1. (Critical path) delay, area, and ADP values of 5MR, 7MR, 9MR, 3-of-5 DMMR, 3-of-6 DMMR and 3-of-7 DMMR circuits corresponding to different Xilinx FPGA families

| Redundancy specification | Delay (ns) | Area (BELs) | ADP value |
|---|---|---|---|
| *Spartan 3E (90nm CMOS)* *Type – Commercial* | | | |
| 5MR | 13.056 | 187 | 2441.472 |
| 7MR | 16.550 | 327 | 5411.85 |
| 9MR | 16.493 | 460 | 7586.78 |
| 3-of-5 DMMR | 13.794 | 179 | 2469.126 |
| 3-of-6 DMMR | 13.949 | 211 | 2943.239 |
| 3-of-7 DMMR | 13.721 | 246 | 3375.366 |
| *Virtex 5 (65nm CMOS process)* *Type – Commercial* | | | |
| 5MR | 6.953 | 107 | 743.971 |
| 7MR | 7.149 | 162 | 1158.138 |
| 9MR | 7.990 | 248 | 1981.52 |
| 3-of-5 DMMR | 7.352 | 109 | 801.368 |
| 3-of-6 DMMR | 6.957 | 129 | 897.453 |
| 3-of-7 DMMR | 7.999 | 154 | 1231.846 |
| *QPro Virtex 2 (150nm CMOS process)* *Type: Radiation tolerant* | | | |
| 5MR | 16.156 | 187 | 3021.172 |
| 7MR | 20.056 | 327 | 6558.312 |
| 9MR | 20.208 | 460 | 9295.68 |
| 3-of-5 DMMR | 16.701 | 179 | 2989.479 |
| 3-of-6 DMMR | 16.712 | 211 | 3526.232 |
| 3-of-7 DMMR | 16.504 | 246 | 4059.984 |
| *QPro Virtex E (180nm CMOS process)* *Type: Military grade* | | | |
| 5MR | 18.937 | 187 | 3541.219 |
| 7MR | 23.901 | 327 | 7815.627 |
| 9MR | 25.934 | 478 | 12396.452 |
| 3-of-5 DMMR | 19.713 | 179 | 3528.627 |
| 3-of-6 DMMR | 20.195 | 211 | 4261.145 |
| 3-of-7 DMMR | 20.030 | 246 | 4927.38 |

The 5MR and 3-of-5 DMMR circuits feature the same number of function modules (i.e. five). Hence the area occupied by their function modules would be the same while there may be minor variations in the area occupancies of their respective voter circuits. Since the critical path delay is the summation of maximum propagation delays encountered in the function module and the voter and interconnects, the data path delay encountered in the function module would be roughly constant while the propagation delays encountered in the respective voters and in the interconnect of 5MR and 3-of-5 DMMR circuits may slightly differ. Hence there is likely to be only a slight difference in the area and delay parameters of the 5MR and 3-of-5 DMMR circuits as seen in Table 1.

When considering the ADP values of different redundant circuits it can be seen from Table 1 that in the case of the commercial FPGAs, the ADPs of 5MR circuits are quite lower than the ADPs of 3-of-5 DMMR circuits while in the case of the radiation-tolerant and military grade FPGAs, the ADPs of the 3-of-5 DMMR circuits are quite lower than the ADPs of the 5MR circuits. Overall, across the four FPGAs considered, the 5MR and 3-of-5 DMMR circuits feature quite similar ADP values. When increases in redundancy orders are considered for the NMR and (3-of-M) DMMR designs it can be seen in Table 1 that the latter significantly outperforms the former in terms of the ADP without compromising on the fault tolerance. The main reason for this is the requirement of less number of function modules for the (3-of-M) DMMR scheme compared to the NMR scheme and partly because of the reduced logic complexities of higher order (3-of-M) DMMR voters compared to the respective higher order NMR voter circuits.

Referring to Table 1, it can be noted that with respect to the two commercial FPGAs considered for physical realization, the 3-of-6 DMMR circuit achieves average reduction in ADP by 34.1% compared to the 7MR circuit. With respect to the radiation-tolerant and military grade FPGAs considered for physical realization, the 3-of-6 DMMR circuit achieves average reduction in ADP by about 46% compared to the 7MR circuit. Again referring to Table 1, with respect to the two commercial FPGAs considered for physical realization, the 3-of-7 DMMR circuit achieves average reduction in ADP by 46.7% compared to the 9MR circuit. With respect to the radiation-tolerant and military grade FPGAs considered for physical synthesis, the 3-of-7 DMMR circuit achieves average reduction in ADP by 58.3% than the 9MR circuit.

## 4 Conclusion

In the era of nanoelectronics, reliability and fault tolerance of circuits and systems assumes increasing importance due to several complex technological issues such as random dopant fluctuations, high heat flux, electro-migration, hot carrier effects, negative bias temperature stability, stress-induced variation, electrostatic discharge, process-induced defects, and metrology and other manufacturing defects.

Redundancy is usually implicit in the design of mission and safety-critical electronic circuits and systems to successfully overcome any unexpected fault(s) or failure(s) which might occur during the normal operation. In this context, the NMR scheme is a well-known method for implementing redundant circuits and systems. However, with multiple faults increasingly likely to become commonplace in nanoelectronic circuits and systems, NMR is not considered to be efficient to implement higher levels of redundancy entirely or selectively in a mission or safety-critical circuit or system design due to the exaggerated increases in design weight, cost and design metrics. Given this, the DMMR scheme forms a good alternative. Without compromising on the fault tolerance, the DMMR scheme is able to facilitate reductions in design cost and weight and also could optimize the design metrics better compared to the NMR scheme due to the requirement of less number of function modules and less complex voters for implementing higher levels of redundancy.

This article has considered the FPGA based realizations of example NMR circuits and their counterpart DMMR circuits. In particular, two commercial FPGAs, a radiation-tolerant FPGA and a military grade FPGA were considered as the implementation platforms. In order to achieve maximum fault tolerances of 2, 3 and 4 function modules, the 5MR, 7MR and 9MR circuits corresponding to the generic NMR scheme and their respective redundant counterparts viz. 3-of-5 DMMR, 3-of-6 DMMR and 3-of-7 DMMR circuits corresponding to the generic (3-of-M) DMMR scheme were considered for physical realization. The 5MR and 3-of-5 DMMR circuits have almost the same design metrics since both these require the same number of function modules. However it was observed that the reductions in identical function modules in the case of the 3-of-6 DMMR and 3-of-7 DMMR circuits compared to the 7MR and 9MR circuits, and the reduced logic complexities of 3-of-6 DMMR and 3-of-7 DMMR voters compared to the counterpart 7MR and 9MR voter circuits translated into significant reductions in ADP for the former compared to the latter.

Across the four different FPGA families considered for physical implementation, the 3-of-6 DMMR circuits feature average reduction in ADP by 44.5% compared to the 7MR circuits, and the 3-of-7 DMMR circuits exhibit average reduction in ADP by 56.5% compared to the 9MR circuits. Hence the important inference from this research work is that to implement higher levels of redundancy in FPGA-based mission and safety-critical circuits and systems entirely or selectively to achieve enhanced fault tolerance, the DMMR scheme forms an efficient alternative to the NMR scheme whilst being able to optimize the design metrics, weight and cost.